\documentclass[aps,prr,twocolumn,superscriptaddress,nofootinbib]{revtex4}

\usepackage{graphicx}
\usepackage{dcolumn}
\usepackage{bm}
\usepackage{amssymb}
\usepackage{amsfonts}
\usepackage{latexsym}
\usepackage{enumerate}
\usepackage{color} 
\usepackage{amsmath}
\usepackage[dvipdfmx]{epsfig}
\usepackage{times}
\usepackage{algorithm}
\usepackage{algorithmic}
\usepackage{cases}
\usepackage{amsthm}
\usepackage{fancybox}

\newtheorem{theorem}{Theorem}
\newtheorem{lemma}{Lemma}

\begin{document}
\widetext
\title{Simplest fidelity-estimation method for graph states with depolarizing noise}
\author{Tomonori Tanizawa}
\affiliation{Department of Physics, Chuo University, 1-13-27 Kasuga, Bunkyo-ku, Tokyo 112-8551, Japan}
\author{Yuki Takeuchi}
\email{yuki.takeuchi@ntt.com}
\affiliation{NTT Communication Science Laboratories, NTT Corporation, 3-1 Morinosato Wakamiya, Atsugi, Kanagawa 243-0198, Japan}
\author{Shion Yamashika}
\affiliation{Department of Physics, Chuo University, 1-13-27 Kasuga, Bunkyo-ku, Tokyo 112-8551, Japan}
\author{Ryosuke Yoshii}
\affiliation{Center for Liberal Arts and Sciences, Sanyo-Onoda City University, 1-1-1 Daigaku-Dori, Sanyo-Onoda, Yamaguchi 756-0884, Japan}
\author{Shunji Tsuchiya}
\email{tsuchiya@phys.chuo-u.ac.jp}
\affiliation{Department of Physics, Chuo University, 1-13-27 Kasuga, Bunkyo-ku, Tokyo 112-8551, Japan}

\begin{abstract}
Graph states are entangled states useful for several quantum information processing tasks such as measurement-based quantum computation and quantum metrology.
As the size of graph states realized in experiments increases, it becomes more essential to devise efficient methods estimating the fidelity between the ideal graph state and an experimentally-realized actual state.
Any efficient fidelity-estimation method, in general, must use multiple experimental settings, i.e., needs to switch between at least two measurements.
Recently, it has been shown that a single measurement is sufficient if the noise can be modeled as the phase-flip error.
Since the bit-flip error should also occur in several experiments, it is desired to extend this simplest method to noise models that include phase and bit-flip errors.
However, it seems to be nontrivial because their result strongly depends on properties of the phase-flip error.
In this paper, by analyzing effects of the bit-flip error on stabilizer operators of graph states, we achieve the extension to the depolarizing noise, which is a major noise model including phase and bit-flip errors.
We also numerically evaluate our simplest method for noise models interpolating between the phase-flip and depolarizing noises.
\end{abstract}
\maketitle

\section{Introduction}
\label{I}
Graph states~\cite{BR01} are entangled states useful for several quantum information processing tasks such as measurement-based quantum computation (MBQC)~\cite{RB01}, quantum metrology~\cite{SM20}, and quantum communication~\cite{ATL15}.
Given this versatility, tremendous theoretical~\cite{SMSN12,BHKM12,MRRBMDK14,ITTIY14,GSBR15} and experimental~\cite{LZGGZYGYP07,TKYKI08,YWXLPBPLCP12,YWCGFRCLLDCP12,WCLHLCLSWLLHJPLLCLP16,WLHCSLCLFJZLLLP18,WLYZ18,GCZWZDYRWLCZLLXGSCWPLZP19,MHH19} efforts have been devoted to increase the size of graph states.
As the size $n$ of graph states realized in experiments increases, it becomes more essential to devise efficient methods estimating the fidelity $\langle G|\rho|G\rangle$ between the ideal $n$-qubit graph state $|G\rangle$ and an experimentally-realized actual state $\rho\equiv\mathcal{E}(|G\rangle\langle G|)$ that suffers from some noise $\mathcal{E}$.
This fidelity estimation is also called the verification of graph states.
So far, several efficient verification methods have been proposed for graph states~\cite{HM15,MNS16,FH17,PLM18,HH18,TM18,TMMMF19,ZH19,ZH19A,MK20,DHZ20,LZH23}.
These methods proceed as follows: (i) Each qubit of $N_c$ copies of $\rho$ is given to a verifier one by one.
(ii) The verifier randomly chooses a measurement basis from $N_m$ kinds of measurements and measures the received state $\rho$ in this basis.
He/she repeats the same procedure for all the copies of $\rho$.
(iii) By processing all measurement outcomes with a classical computer, he/she outputs an estimated value of (or a lower bound on) the fidelity $\langle G|\rho|G\rangle$.
In most cases, to reduce the burden on the verifier as much as possible, only non-adaptive single-qubit projective measurements and efficient classical operations are required for the verifier.
In this paper, we consider the same restriction on the verifier.

In the evaluation of verification protocols, two parameters $N_c$ and $N_m$ are usually considered.
So far, several attempts to reduce the number $N_c$ of copies have been done, and Zhu and Hayashi have finally constructed an optimal verification protocol~\cite{ZH19A} such that $N_c=\Theta(\epsilon^{-1}\log{\delta^{-1}})$ to guarantee $\langle G|\rho|G\rangle\ge 1-\epsilon$ with significance level $\delta$.
As a remarkable property, the number $N_c$ in their optimal protocol does not depend on the size $n$ of the graph state $|G\rangle$.

On the other hand, the optimality of the number $N_m$ of measurement settings is less explored.
In many practical cases, the switching of measurement settings could be slow, and in some cases, it may be demanding or impossible (e.g., see Ref.~\cite{LZLZ21}).
Furthermore, since the measurement error is the most dominant in some state-of-the-art experiments~\cite{google}, the reduction of the number of measurement settings should be helpful to realize verification protocols with high accuracy.
Therefore, it is important to reduce $N_m$ (ultimately to one) under the assumption that the verifier can perform only non-adaptive single-qubit projective measurements.
However, under this assumption, it has been shown that at least two measurement settings are required for the verification of any bipartite pure entangled state if $\mathcal{E}$ is an arbitrary noise~\cite{LZLZ21}.
Since bipartite pure entangled states include a subclass of graph states, their result prevents the possibility of $N_m=1$ for general noises.
Even if adaptive measurements are allowed for the verifier, at least two measurement settings are still necessary~\cite{ZH19A}.
Indeed, although several verification protocols were proposed for graph states~\cite{HM15,MNS16,FH17,PLM18,HH18,TM18,TMMMF19,ZH19,ZH19A,MK20,DHZ20,LZH23}, they require multiple measurement bases.

Recently, by restricting the noise model (i.e., by fixing $\mathcal{E}$), a verification protocol achieving $N_m=1$ has been constructed~\cite{ATYT22}.
In this protocol, $\mathcal{E}$ is assumed to be the phase-flip error, and they have achieved the optimal number of $N_m$ by using properties of the phase-flip error.
More precisely, from the commutation relations between Pauli operators and the phase-flip error, they have shown that measurements of a single stabilizer operator of $|G\rangle$ are sufficient to estimate a lower bound on the fidelity with high accuracy.
Therefore, it is nontrivial whether $N_m=1$ can be achieved for other noise models including bit-flip errors.

In this paper, we propose a verification protocol achieving $N_m=1$ for graph states in the presence of the depolarizing noise (see also Eq.~(\ref{dplns})).
Since the depolarizing noise is a major noise model used in several theoretical analyses of quantum error correction~\cite{RHG07,WFH11,BAOKM12} and error mitigation~\cite{TEMG22,TTG22,TSY22}, our protocol should also be compatible with other methods handling errors.
To construct our protocol, we analyze the effect of depolarizing noise on the fidelity $\langle G|\rho|G\rangle$.
As a well-known fact, the depolarizing noise on $n$ qubits can be written as a classical mixture of Pauli noises~\cite{NC00}.
We observe that Pauli noises definitely reduce the fidelity if and only if they do not coincide with any stabilizer operator of $|G\rangle$. 
From this observation, we obtain a single measurement from which we can obtain an approximate value of the fidelity. 
By using this measurement, we propose a verification protocol for graph states with the depolarizing noise that satisfies $N_c=\Theta(\epsilon^{-2}\log{\delta^{-1}})$ and $N_m=1$.
As concrete applications, we apply our verification protocol to $n$-qubit fully-connected graph states, which can be converted to $n$-qubit Greenberger-Horne-Zeilinger (GHZ) states by local Clifford operations~\cite{VDM04}, and cluster states. 
Since cluster states are resource states of universal MBQC, and GHZ states can be used to perform quantum sensing achieving the Heisenberg limit~\cite{TA14} and non-adaptive MBQC with linear side-processing (${\rm NMQC}_\oplus$)~\cite{HCLB11}, our protocol can be used to make these protocols verifiable.

We also evaluate our protocol for noise models other than the depolarizing noise. 
First, we consider the noise model where the phase-flip or depolarizing noise is randomly applied.
We show that although it is unknown which noise is applied, our protocol works well for some cluster states.
Then we consider noise models interpolating between the phase-flip and depolarizing noises.
We numerically evaluate how well our protocol works for these noise models.
Lastly, we compare our protocol with previous protocols.

The rest of this paper is organized as follows: in Sec.~\ref{II}, we introduce graph states and the depolarizing noise.
In Sec.~\ref{III}, we propose our verification protocol using only a single stabilizer measurement.
In Sec.~\ref{IV}, as concrete examples, we apply our protocol in Sec.~\ref{III} to fully-connected graph states and cluster states.
We also evaluate our protocol in noise models other than the depolarizing noise.
In Sec.~\ref{V}, we compare our protocol with previous protocols.
Section~\ref{VI} is devoted to the conclusion and discussion.
In Appendices A, B, C, and D, we give a proof of Lemma~\ref{lemma1}, derivation of Eq.~(\ref{stblzr_general}), proof of Theorem~\ref{theorem1}, and derivation of Eq.~(\ref{uprbnd_GHZ}), respectively.

\medskip
\section{Graph states in the depolarizing channel}
\label{II}
In this section, we introduce graph states in the depolarizing channel.
To this end, we first define graph states~\cite{BR01}.
A graph $G\equiv(V,E)$ is a pair of the set $V$ of $n$ vertices and the set $E$ of edges.
The $n$-qubit graph state $|G\rangle$ that corresponds to the graph $G$ is defined as
\begin{eqnarray}
\label{graph}
|G\rangle\equiv\left(\prod_{(i,j)\in E}CZ_{i,j}\right)|+\rangle^{\otimes n},
\end{eqnarray}
where $|+\rangle\equiv(|0\rangle+|1\rangle)/\sqrt{2}$ with $|0\rangle$ and $|1\rangle$ being, respectively, eigenstates of the Pauli-$Z$ operator with eigenvalues $+1$ and $-1$, and $CZ_{i,j}$ is the controlled-$Z$ ($CZ$) gate applied on the $i$th and $j$th qubits.
The stabilizer generators $\{g_i\}_{i=1}^n$ for $|G\rangle$ are defined as
\begin{eqnarray}
\label{gstabilizer}
g_i\equiv X_i\left(\prod_{j:\ (i,j)\in E}Z_j\right).
\end{eqnarray} 
Here, $X_i$ and $Z_j$ are the Pauli-$X$ and $Z$ operators for the $i$th and $j$th qubits, respectively, and the product of $Z_j$ is taken over all vertices $j$ such that $(i,j)\in E$.
For any $i$ and $j$, two stabilizer generators commute, i.e., $[g_i,g_j]=0$.
The graph state $|G\rangle$ is the unique common eigenstate of $\{g_i\}_{i=1}^n$ with eigenvalue $+1$ , i.e., $g_i|G\rangle=|G\rangle$ for any $i$.

A stabilizer $S_{\ell}$ is a product of stabilizer generators such that $S_{\ell}\equiv\prod_{i=1}^ng_i^{\ell_i}$, where ${\ell}\equiv \ell_1\ell_2\ldots \ell_n\in\{0,1\}^n$.
It is a tensor product of $n$ single-qubit operators. More precisely, by using $s\in\{0,1\}$ and $\tau_i\in\{I,X,Y,Z\}$, where $I$ and $Y=iXZ$ are the two-dimensional identity operator and Pauli-$Y$ operator, respectively, it can be written as
\begin{equation}
    S_{\ell}=(-1)^s\bigotimes_{i=1}^n\tau_i.
    \label{stabilizer}
\end{equation}
For any $\ell$, the equality $S_{\ell}|G\rangle=|G\rangle$ can be easily checked from Eqs.~(\ref{graph}) and (\ref{gstabilizer}).

The depolarizing channel is represented by the superoperator \cite{NC00}
\begin{eqnarray}
\mathcal{E}(\cdot)\equiv(1-p)I(\cdot)I+\frac{p}{3}\left [X(\cdot)X+Y(\cdot)Y+Z(\cdot)Z\right ].
\label{dplns}
\end{eqnarray}
It operates independently on each qubit, where bit-flip (Pauli-$X$ error), phase-flip (Pauli-$Z$ error), and bit-phase-flip errors (Pauli-$Y$ error) occur with equal probability $p/3$.
The depolarizing channel is a major noise model that is used in several analyses as explained in Sec.~\ref{I}.

Let $\ovalbox{$|\psi\rangle$}\equiv|\psi\rangle\langle\psi|$ for any pure state $|\psi\rangle$.
The density operator $\rho\equiv\mathcal{E}^{\otimes n}(|G\rangle\langle G|)$ for the graph state in the depolarizing channel can be written as 
\begin{widetext}
\begin{align}
\rho=&(1-p)^n\ovalbox{$|G\rangle$}+(1-p)^{n-1}\frac{p}{3}\sum_{i=1}^n\sum_{\mu=1}^3\ovalbox{$\sigma_{\mu i}|G\rangle$}+(1-p)^{n-2}\left(\frac{p}{3}\right)^2\sum_{1\le i< j\le n}\sum_{\substack{1\le\mu\le 3 \\ 1\le\nu\le3}}\ovalbox{$\sigma_{\mu i}\sigma_{\nu j}|G\rangle$}+\dots\notag\\
=&(1-p)^n\ovalbox{$|G\rangle$}
+\sum_{m=1}^n(1-p)^{n-m}\left(\frac{p}{3}\right)^m\sum_{\substack{1\le i_1<i_2<\ldots<i_m\le n \\ \mu_1\mu_2\ldots\mu_m\in\{1,2,3\}^m}}\ovalbox{$(\prod_{k=1}^m\sigma_{\mu_ki_k})|G\rangle$}~,
\label{dperror}
\end{align}
\end{widetext}
where $\sigma_{1i}\equiv X_i$, $\sigma_{2i}\equiv Y_i$, and $\sigma_{3i}\equiv Z_i$.

We define the fidelity between two states $\rho_1$ and $\rho_2$ as
\begin{eqnarray}
F\equiv\left(\textrm{Tr}\sqrt{\sqrt{\rho_1}\rho_2\sqrt{\rho_1}}\right)^{2}.
\label{fidelty_def}
\end{eqnarray}
Note that the fidelity is defined as $\sqrt{F}$ in Ref.~\cite{NC00}. For ease of our argument, we use the definition in Eq.~(\ref{fidelty_def}).
Therefore, the fidelity between the graph state $\rho$ in the depolarizing channel and the ideal state $|G\rangle\langle G|$ 
can be written as
\begin{align}
F=
&\langle G|\rho|G\rangle\notag\\
=&(1-p)^n+\sum_{m=1}^n(1-p)^{n-m}\left(\frac{p}{3}\right)^m\notag\\
&\times\sum_{\substack{i_1<\ldots<i_m \\ \mu_1\ldots\mu_m}}\langle G|\left(\prod_{k=1}^m\sigma_{\mu_ki_k}\right)|G\rangle^2.
\label{fidelity_graph}
\end{align}
The following lemma is useful for evaluation of Eq.~(\ref{fidelity_graph}).
\begin{lemma}
\label{lemma1}
Suppose $|G\rangle$ is any $n$-qubit graph state, and $\sigma_{\mu i}$ is the $\mu$ component of the Pauli operator for the $i$th qubit. Then, for any natural number $m(\le n)$, 
\begin{equation}
    \langle G|\left(\prod_{k=1}^m\sigma_{\mu_ki_k}\right)|G\rangle^2=1,
\end{equation}
if and only if one of $\pm\prod_{k=1}^m\sigma_{\mu_ki_k}$
coincides with a stabilizer of $|G\rangle$. Otherwise, it vanishes.
\end{lemma}
Although Lemma~\ref{lemma1} can be straightforwardly obtained from basic properties of graph states (e.g., see Ref.~\cite{ACCDDA10}), we give a proof of Lemma \ref{lemma1} in Appendix A for the completeness of our paper.
From Lemma \ref{lemma1},
\begin{eqnarray}
\sum_{\substack{i_1<\ldots<i_m \\ \mu_1\ldots\mu_m}}\langle G|\left(\prod_{k=1}^m\sigma_{\mu_ki_k}\right)|G\rangle^2
\end{eqnarray}
is equal to the number of the stabilizers that are products of $m$ Pauli operators.
Since the number of stabilizers increases exponentially with $n$, in general, it would be hard to derive the exact value of $F$ for large $n$. However, in Sec.~\ref{IVA}, we show that $F$ can be represented by a simple formula (Eq.~(\ref{fulgraph_fidexact})) for fully-connected graph states.

We can assume, without loss of generality, that graph states we consider in this paper have no isolated single qubits, because verification of isolated single qubits can be performed independently of other connected qubits. Each stabilizer of a connected graph state is a product of at least two Pauli operators, so that the first-order error term ($m=1$) in Eq.~(\ref{fidelity_graph}) vanishes.
The second-order error term ($m=2$) also vanishes for any 
graph state that has no stabilizers consisting of two Pauli operators, such as a two-dimensional cluster state with $n>4$.

Using the identity
\begin{equation}
|G\rangle\langle G|=\prod_{i=1}^{n}\frac{g_{i}+I_{i}}{2}=\frac{1}{2^{n}}\sum_{\ell}S_{\ell},
\label{stabilizeridentity}
\end{equation}
the fidelity can also be written as 
\begin{eqnarray}
F=\textrm{Tr}(\rho|G\rangle\langle G|)=\frac{1}{2^{n}}\sum_{\ell}\textrm{Tr}(\rho S_{\ell}).
\label{fidelity_S}
\end{eqnarray}
Equation~(\ref{fidelity_S}) indicates that the fidelity can be estimated exactly from the average of all the stabilizers
$\{S_{\ell}\}_{\ell\in\{0,1\}^n}$. That is, $2^n$ kinds of measurement settings are required.
Note that only a polynomial number of them are chosen uniformly at random and performed in actual experiments.
However, since chosen measurements vary in each experiment, the estimation of the fidelity requires the ability of performing any measurement in the $2^n$ stabilizer measurements.

\medskip
\section{Fidelity estimation by measuring a single stabilizer}
\label{III}
 
In this section, we discuss our idea used in constructing our simplest fidelity-estimation protocol for graph states in the depolarizing channel.
The average of the stabilizer is given as 
\begin{align}
&{\rm Tr}(\rho S_{\ell})=(1-p)^{n}
+\sum_{m=1}^n(1-p)^{n-m}\left(\frac{p}{3}\right)^m\notag\\
&\cdot\left[\sum_{\substack{i_1<\ldots<i_m \\ \mu_1\ldots\mu_m}}\langle G|\left(\prod_{k=1}^m\sigma_{\mu_ki_k}\right)S_\ell\left(\prod_{k=1}^m\sigma_{\mu_ki_k}\right)|G\rangle\right].
\label{stblzr_ave}
\end{align}
Since the first-order error term vanishes in Eq.~(\ref{fidelity_graph}), the fidelity can be well approximated by the average of a stabilizer without the first-order error term when $p\ll 1$.
By comparing Eqs.~(\ref{fidelity_graph}) and (\ref{stblzr_ave}), for $p\ll 1$, we expect that the fidelity can be accurately estimated by measuring a single stabilizer for which the first-order error term vanishes.

Let us investigate the condition on which the first-order error term in Eq.~(\ref{stblzr_ave}) vanishes. 
Recall that $\tau_i$ is a single-qubit operator for the $i$th qubit of $S_{\ell}$. $\tau_i$ commutes with each of $X$, $Y$, and $Z$ in the case of $\tau_i=I$, while it commutes with only one of them and anticommutes with the others in the case of $\tau_i\in\{X,Y,Z\}$. We thus obtain
\begin{equation}
\sum_{\mu=1}^{3}\langle G|\sigma_{\mu i}S_{\ell}\sigma_{\mu i}|G\rangle=
\left\{
\begin{array}{l}
3\quad (\tau_i=I)\\
-1\quad (\tau_i\in\{X,Y,Z\}).
\end{array}
\right.
\end{equation}
Using this relation, we obtain
\begin{align}
\sum_{i=1}^n\sum_{\mu=1}^3\langle G|\sigma_{\mu i}S_{\ell}\sigma_{\mu i}|G\rangle\notag
&=
3n_I-(n-n_I)\\
&=4n_{I}-n,
\label{C(1)}
\end{align}
where $n_I$ denotes the number of $I$ in $S_{\ell}$. Hence, the condition for the first-order error term to vanish is
\begin{equation}
    n_I=\frac{n}{4}.
    \label{condition}
\end{equation}

Generalizing the calculation of the first-order error term, the average of a stabilizer can be calculated in general as
\begin{align}
{\rm Tr}(\rho S_{\ell})&=(1-p)^n\notag\\
&+\sum_{m=1}^n C(m)(1-p)^{n-m}\left(\frac{p}{3}\right)^m,\label{general}
\end{align}
\begin{align}
C(m)&\equiv\sum_{\substack{i_1<\ldots<i_m \\ \mu_1\ldots\mu_m}}\langle G|\left(\prod_{k=1}^m\sigma_{\mu_ki_k}\right)S_\ell\left(\prod_{k=1}^m\sigma_{\mu_ki_k}\right)|G\rangle\notag\\
&=\sum_{j=w(\ell,m,n)}^{f(\ell,m)}(-1)^{m-j}3^{j}\binom{n_{I}}{j}\binom{n-n_{I}}{m-j},
\label{generalC}
\end{align}
where $f(\ell,m)\equiv{\rm min}\{m,n_I\}$ and $w(\ell,m,n)\equiv{\rm max}\{0,m+n_I-n\}$. We have assumed $\binom{0}{0}=1$.

$C(m)$ corresponds to the sum of the averages of $S_{\ell}$ in the case of $m$ errors.  
The term $\binom{n_I}{j}\binom{n-n_I}{m-j}$ expresses the number of cases in which $j$ errors occur among the $n_I$ qubits for which $\tau_i=I$, while the other $(m-j)$ errors occur on $(n-n_I)$ qubits for which $\tau_i\in\{X,Y,Z\}$.

From Eq.~(\ref{generalC}), the coefficient $C(2)$ is given as
\begin{align}
C(2)=8n_I^2-4(n+1)n_I+\frac{n(n-1)}{2}.
\label{secondC}
\end{align}
By comparing the second-order error terms ($m=2$) in Eqs.~(\ref{fidelity_graph}) and (\ref{general}), it should be preferable to set $C(2)$ as a non-negative number for the verification with high accuracy.
Particularly in the case of cluster states, $C(2)=0$ is desirable.
However, in the case that the first-order error term vanishes, substituting $n_I=n/4$ in Eq.~(\ref{secondC}), we find that the second-order error term is negative as $C(2)=-3n/2$.

From Eq.~(\ref{generalC}), one finds that $C(m)$ is equivalent to the coefficient of the term $x^{n-m}y^m$ in the expansion of the polynomial $(x+3y)^{n_I}(x-y)^{n-n_I}$. Then, substituting $x=1-p$ and $y=p/3$ in $(x+3y)^{n_I}(x-y)^{n-n_I}$, we finally obtain a simple analytical expression for the average of the stabilizer as follows:
\begin{eqnarray}
{\rm Tr}(\rho S_{\ell})
&=&\left[(1-p)+3\cdot\frac{p}{3}\right]^{n_I}\left[(1-p)-\frac{p}{3}\right]^{n-n_I}
\label{generallong}\\
&=&\left(1-\frac{4}{3}p\right)^{n-n_I}.
\label{generalshort}
\end{eqnarray}
It is clear from Eq.~(\ref{generalshort}) that for fixed $n$ and $p$, ${\rm Tr}(\rho S_{\ell})$ is determined solely by the number of $I$ in $S_{\ell}$. The average of the stabilizer decreases from unity as $p$ increases in Eq.~(\ref{generalshort}), because eigenstates of $S_{\ell}$ with eigenvalue $-1$ are mixed to the pure graph state $|G\rangle\langle G|$ by the depolarizing noise. ${\rm Tr}(\rho S_{\ell})$ becomes negative when $p>3/4$ in the case of odd $n-n_I$.

Equation~(\ref{generalshort}) can be derived using an alternative approach, recognizing the tensor structure of both the stabilizer operator and the depolarizing noise.  Using this approach, the average of a stabilizer under the general noise model
\begin{align}
\mathcal{E}(\cdot)=&(1-p_x-p_y-p_z)(\cdot)\notag\\
&+\left[p_xX(\cdot)X+p_yY(\cdot)Y+p_zZ(\cdot)Z\right]
\label{generalnoise}
\end{align}
can be derived as
\begin{align}
{\rm Tr}(\rho S_{\ell})
=&
\left(1-2p_y-2p_z\right)^{n_X}\left(1-2p_z-2p_x\right)^{n_Y}\notag\\
&\cdot\left(1-2p_x-2p_y\right)^{n_Z}.
\label{stblzr_general}
\end{align}
Here, $n_X$, $n_Y$ and $n_Z$ are the numbers of $X$, $Y$, and $Z$ in the stabilizer $S_{\ell}$, respectively. Equation~(\ref{stblzr_general}) reduces to Eq.~(\ref{generalshort}) when $p_x=p_y=p_z=p/3$. We provide a detailed description of the derivation of Eq.~(\ref{stblzr_general}) in Appendix B.

Equations~(\ref{fidelity_S}) and (\ref{generalshort}) lead to a simple expression for the fidelity
\begin{equation}
F=\frac{1}{2^n}\sum_{\ell\in\{0,1\}^n}\left(1-\frac{4}{3}p\right)^{n-n_I(\ell)},
\label{fidelitygene}
\end{equation}
where $n_I(\ell)$ denotes the number of $I$ in the stabilizer $S_{\bf \ell}$.

We have shown that $C(1)=0$ when $n_I=n/4$ in Eq.~(\ref{C(1)}). In fact, setting $n_I=n/4$ in Eqs.~(\ref{generallong}) and (\ref{generalshort}), we obtain  
\begin{align}
{\rm Tr}(\rho S_{\bm\ell})
=&\left(1-\frac{4}{3}p\right)^{3n/4}\notag\\
=&\left[(1-p)^4-\cfrac{2}{3}~p^2(1-p)^2\right.\notag\\
&\left.+\cfrac{8}{27}~p^3(1-p)-\cfrac{1}{27}~p^4\right]^{n/4}.
\label{expectation}
\end{align}
Expanding the final expression and comparing with Eq. \eqref{fidelity_graph}, it is clear that there is no first-order error term ($m=1$) in Eq.~(\ref{expectation}).

From this observation, we expect that the fidelity can be well estimated by measuring a single stabilizer that satisfies the condition Eq.~(\ref{condition}).
In fact, the following theorem holds.
\begin{theorem}
\label{theorem1}
Let $|G\rangle$ be an $n$-qubit ideal graph state with $n=4k$ for some natural number $k$.
Let ${\mathcal A}$ be the set of stabilizers $S$ of $|G\rangle$ such that $S=(-1)^s\otimes_{i=1}^n\tau_i$, where $\tau_i\in\{X,Y,Z\}$ for $3k$ kinds of $i$'s, $\tau_i=I$ for other $i$'s, and $s\in\{0,1\}$.
Let $F\equiv\langle G|\rho|G\rangle$ be the fidelity between $|G\rangle$ and an $n$-qubit graph state $\rho$ (Eq.~(\ref{dperror})) in the depolarizing channel with the error probability $p$. The fidelity $\tilde F=(1-p)^{4k}$ up to the first-order error can be approximated by the expectation value $F_{\rm est}\equiv{\rm Tr}(\rho S)=(1-4p/3)^{3k}$ of any single stabilizer $S$ in the set ${\mathcal A}$, such that
\begin{eqnarray}
0\le \tilde F-F_{\rm est}<\frac{2}{3k}
\label{lwrbnd}
\end{eqnarray}
for $0\le p\le 3/4$.
\end{theorem}

\begin{figure}[t]
\includegraphics[width=8cm,clip]{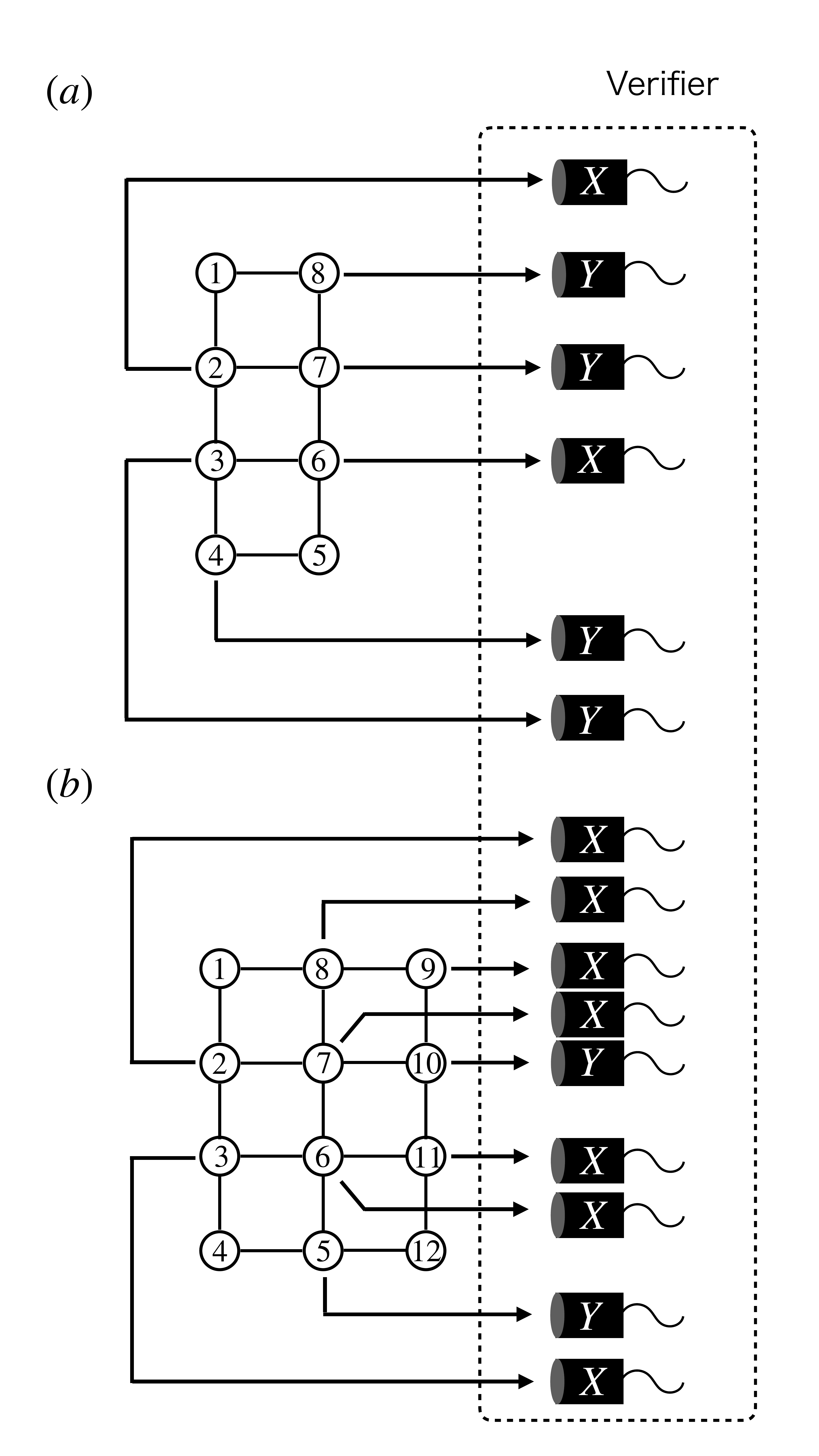}
\caption{Schematic diagram of our simplest verification protocol.
A quantum computer generates graph states $\rho$ in the depolarizing channel and sends each qubit one by one.
A verifier just measures a single stabilizer $S_{\ell}$, which has $3n/4$ Pauli operators, for each received state $\rho$ by using only single-qubit Pauli measurements.
No quantum memory is required for the verifier.}
\label{dverification}
\end{figure}

Figures~\ref{dverification} (a) and (b) illustrate examples of the stabilizers $S_{\ell}$ in the set $\mathcal A$ for two-dimensional cluster states. They satisfy the condition ${\rm wt}(\ell)\equiv \sum_{i=1}^n \ell_i=3k$ and $\tau_i=I$ for arbitrary qubit $i$ for which $\ell_i=0$. Namely, $S_{\ell}$ consists of $3k$ generators, and arbitrary qubit unoccupied by a generator is connected to occupied qubits with an even number of edges. Another example of such stabilizers is given in Sec.~\ref{IV} for the fully-connected graph states.

Our main contribution is to derive Eq.~(\ref{generalshort}) from which 
Theorem~\ref{theorem1} can be immediately obtained.
A rigorous proof of Theorem~\ref{theorem1} is given in Appendix C. Note that Theorem 1 holds for any graph state with $n=4k$ under the condition that the graph state has no isolated single qubits, and the set $\mathcal{A}$ is not empty.
Theorem~\ref{theorem1} implies that the more the number $n$ of qubits increases, the more the precision of the estimation for $\tilde F$ by $F_{\rm est}$ improves.
When the second-order error term in Eq.~(\ref{fidelity_graph}) vanishes, the fidelity up to the second-order error is also $\tilde F$, and so the estimation of $F$ by $F_{\rm est}$ improves further. This is the case for two-dimensional (2D) cluster states, as we demonstrate it in the next section.

Based on Theorem~\ref{theorem1}, our verification protocol runs as follows:
\begin{enumerate}
\item A quantum computer generates $N$ graph states $\rho^{\otimes N}$ in the depolarizing channel and sends them to a verifier.
\item The verifier measures $S_{\ell}$ that satisfies the condition in Eq.~(\ref{condition}) on each received state $\rho$.
\item The verifier outputs 
\begin{eqnarray}
\label{estimate}
\tilde F_{\rm est}\equiv\cfrac{\sum_{i=1}^No_i}{N}
\end{eqnarray}
as an estimated value of the fidelity, where $o_i\in\{+1,-1\}$ denotes the $i$th outcome for $1\le i\le N$.
\end{enumerate}

$\tilde F_{\rm est}$ converges to $F_{\rm est}$ in the limit of large $N$. In fact, the Hoeffding inequality~\cite{H63} guarantees that when $N=\lceil2/\epsilon^2\log{(2/\delta)}\rceil$, the inequality
\begin{eqnarray}
\label{Hoeffding}
\left|F_{\rm est}-\tilde F_{\rm est}\right|\le\epsilon
\end{eqnarray}
holds with probability at least $1-\delta$.
Here, $\lceil\cdot\rceil$ is the ceiling function.

The measurement of $S_{\ell}$ in step 2 can be realized by single-qubit Pauli measurements because $S_{\ell}$ is a tensor product of Pauli operators.
Furthermore, by sequentially sending qubits one by one in step 1, no quantum memory is required for the verifier.
To illustrate our protocol, we give concrete examples in Fig.~\ref{dverification}.

While the constraint in our protocol on the number of qubits, which must be a multiple of 4, may seem restrictive, we point out that it does not limit its practical applications.
The central goal of our protocol is to verify high fidelity for graph states.
Once high fidelity has been established for a graph state using our protocol, the number of qubits of the graph state can be changed by measurements in the $Z$ basis because measuring a qubit in the $Z$ basis results in breaking the edges and disconnecting the qubit from the graph.

\medskip
\section{Applications}
\label{IV}

In this section, we first discuss the estimation of the fidelity for the fully-connected graph states and the cluster states.
Particularly, the cluster states are important resource states for MBQC, which allows universal quantum computation.
Theorem~\ref{theorem1} just guarantees that our simplest verification protocol outputs the estimated value that is close to the true value $F$ of the fidelity only when $p$ is sufficiently small.
We numerically show that $F_{\rm est}$ becomes precise approximations for any $0\le p\le 1/2$ in the case of fully-connected graph and cluster states.
Then, we evaluate our protocol for noise models other than the depolarizing noise.

\subsection{Fully-connected graph states}
\label{IVA}

\begin{figure}[t]
\includegraphics[width=8cm,clip]{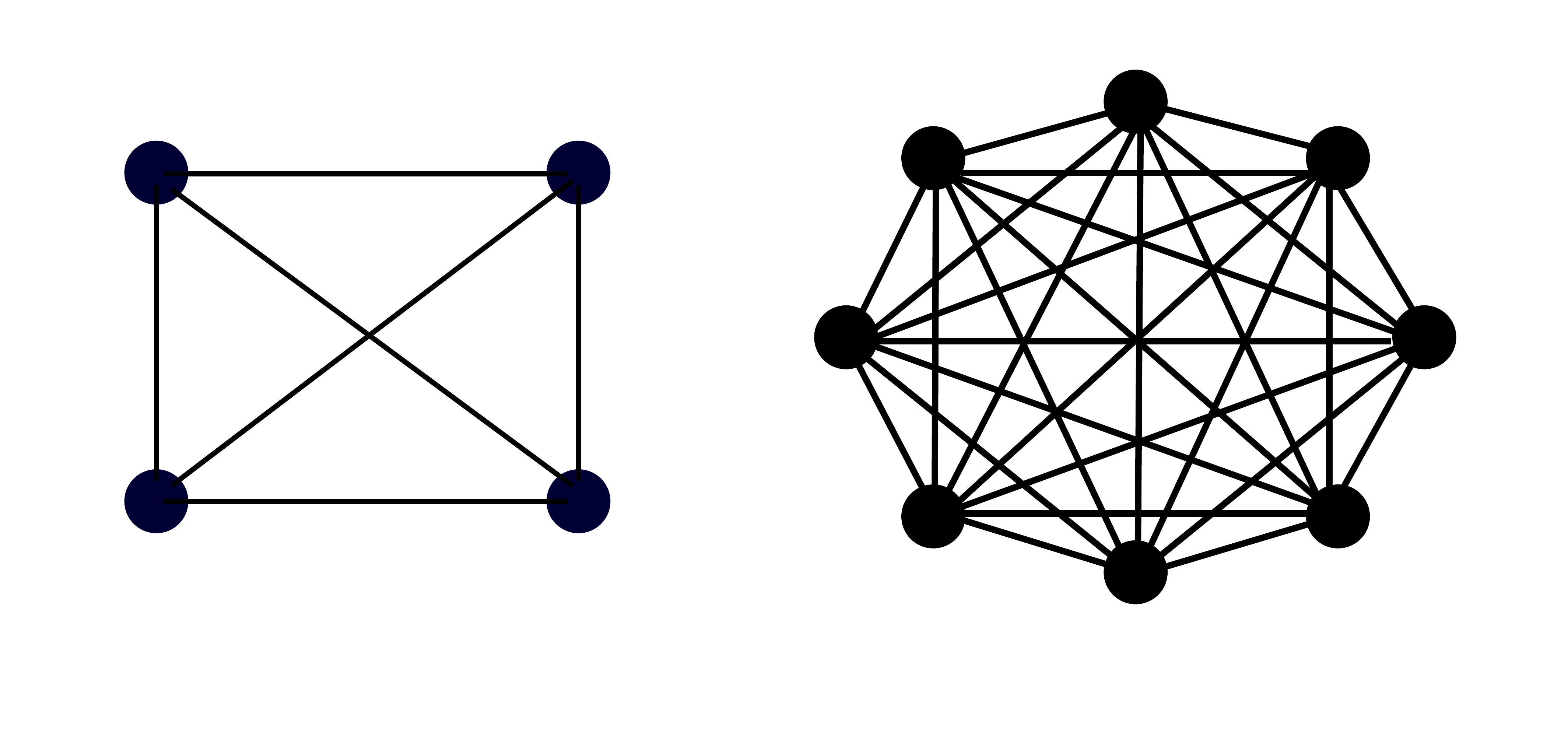}
\caption{Fully-connected graphs with four (left) and eight vertices (right). The black dots and the solid lines are vertices and edges, respectively.}
\label{fulgraph}
\end{figure}

We consider fully-connected graphs, in which each of the vertices is connected with all the other vertices by the edges, as shown in Fig.~\ref{fulgraph}.
Before applying Theorem 1 to fully-connected graph states with $n=4k$ ($k\in \mathbb N$) qubits, 
we first evaluate the fidelity of any fully-connected graph state.
According to Eq.~(\ref{fidelitygene}), since $n_I(\ell)$ for all $\ell$'s are required to evaluate the fidelity, it should not be easy to derive it in general.
For the fully-connected graph states, however, Eq.~(\ref{fidelitygene}) can be easily evaluated.

In the case of even ${\rm wt}(\ell)=\sum_{i=1}^n\ell_i$, i.e., $S_{\ell}$ is a product of even $g_i$'s, we obtain $n_I({\ell})=n-{\rm wt}(\ell)$, because $\tau_i=I$ for $\ell_i=0$, and $\tau_i=X$ or $Y$ for $\ell_i=1$. Meanwhile, $n_I=0$ in the case of odd ${\rm wt}(\ell)$, because $\tau_i=Z$ for $\ell_i=0$, and $\tau_i=X$ or $Y$ for $\ell_i=1$. We thus obtain
\begin{align}
&2^nF=\sum_{\rm wt:\ even}\binom{n}{\rm wt}\left(1-\frac{4}{3}p\right)^{\rm wt}\notag\\
&\quad+\sum_{\rm wt:\ odd}\binom{n}{\rm wt}\left(1-\frac{4}{3}p\right)^n\notag\\
&=2^{n-1}\left[\left(1-\frac{2}{3}p\right)^n+\left(\frac{2}{3}p\right)^n+\left(1-\frac{4}{3}p\right)^n\right].
\label{fulgraph_fidexact}
\end{align}
Here, we have used the following relations
\begin{align}
&\sum_{j:\ {\rm even}}\binom{n}{j}x^j=\frac{1}{2}\left[(1+x)^n+(1-x)^n\right],\\
&\sum_{j:\ {\rm odd}}\binom{n}{j}=2^{n-1}.
\label{sumodd}
\end{align}

Now we discuss the estimation of the fidelity for fully-connected graph states with $n=4k$ qubits. In the case of even $k$, any stabilizer $S_{\ell}$ with ${\rm wt}=3k$ satisfies the condition $n_I=n/4=k$, because $\tau_i=I$ for $\ell_i=0$, and $\tau_i=X$ or $Y$ for $\ell_i=1$.
Meanwhile, in the case of odd $k$, any fully-connected graph state with $n=4k$ qubits has no stabilizers that satisfy the condition $n_I=n/4=k$, because any stabilizer $S_{\ell}$ with ${\rm wt}=3k$ has $\tau_i=Z$ for $\ell_i=0$, and $\tau_i=X$ or $Y$ for $\ell_i=1$. Theorem \ref{theorem1} can be thus applied to any fully-connected graph state with $n=8k$ ($k\in \mathbb N$) qubits.

\begin{figure}[t]
\includegraphics[width=8cm,clip]{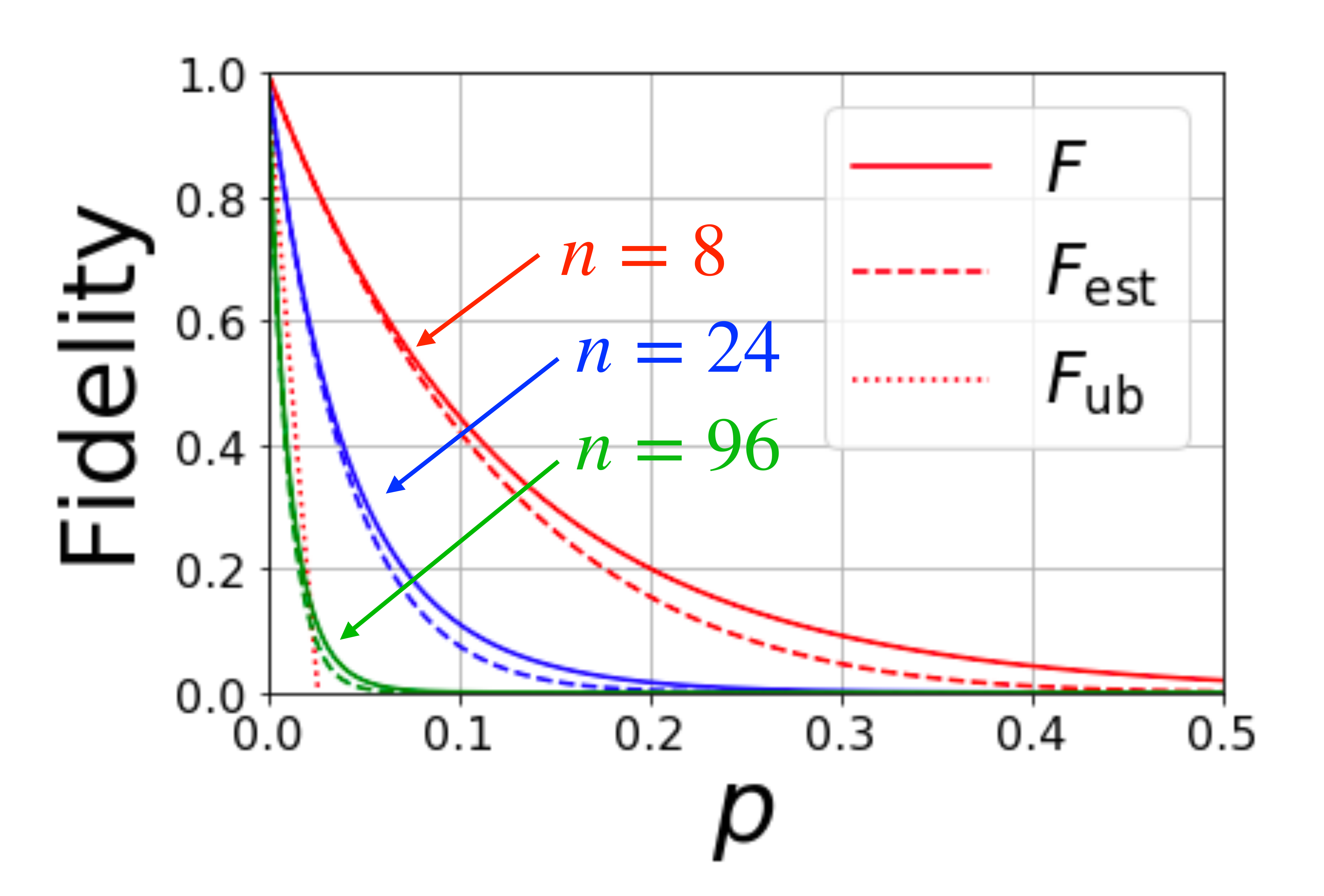}
\caption{Comparison of $F$ (solid lines) and $F_{\rm est}$ (dashed lines) as functions of the error probability $p$ for the fully-connected graph states with $n=8$, $24$, and $96$ qubits. The dotted line represents $F_{\rm ub}$ in Eq.~(\ref{Fubfully}) with $n=8$.}
\label{fulgraph_fidelity}
\end{figure}

Figure~\ref{fulgraph_fidelity} shows the comparison of $F$ in Eq.~(\ref{fulgraph_fidexact}) and $F_{\rm est}=(1-4p/3)^{6k}$ for the fully-connected graph states with $n=8k$ qubits. It demonstrates that the estimation of the fidelity $F$ by $F_{\rm est}$ improves as $n$ increases. 
The upper bound on $F-F_{\rm est}$ for $0\leq p\leq 3/4$ can be derived as
\begin{align}
0\leq F-F_{\rm est}<& \frac{1}{2}\left(1-\frac{1}{8k}\right)\left(1-\frac{1}{4k}\right)^{8k-2}\notag\\
&+\frac{1}{2}\left(\frac{2}{3}\right)^{8k}+\frac{1}{3k}.
\label{uprbnd_GHZ}
\end{align}
The derivation is given in Appendix D. The upper bound in Eq.~(\ref{uprbnd_GHZ}) monotonically decreases as $k$ increases and converges to $1/(2e^2)$ in the limit of $k\to \infty$.

The second-order error term $F^{(2)}$ in Eq.~(\ref{fidelity_graph}) is nonzero for the fully-connected graph states. Since any stabilizer with $\rm wt=2$ consists of two Pauli operators, it can be written as
\begin{equation}
    F^{(2)}=\binom{n}{2}(1-p)^{n-2}\left(\frac{p}{3}\right)^2.
\label{F2}
\end{equation}
The relatively large deviation of $F_{\rm est}$ from $F$ in Fig.~\ref{fulgraph_fidelity} reflects the presence of $F^{(2)}$.

In the case of fully-connected graph states, the other error terms can also be derived.
First, from Eq.~(\ref{fidelity_graph}), we obtain
\begin{align}
    F=\sum_{\ell\in\{0,1\}^n}(1-p)^{n_I(\ell)}\left(\frac{p}{3}\right)^{n-n_I(\ell)}.
\label{generalF}
\end{align}
Then, by following the similar argument as used to derive Eq.~(\ref{F2}) and using Eq.~(\ref{sumodd}), we calculate Eq.~(\ref{generalF}) as follows:
\begin{align}
    F=&(1-p)^n+\sum_{k=1}^{ \lfloor n/2\rfloor}\binom{n}{2k}(1-p)^{n-2k}\left(\frac{p}{3}\right)^{2k}\nonumber\\
    &+2^{n-1}\left(\frac{p}{3}\right)^n,
    \label{fulgraph_fidelity2}
\end{align}
where $\lfloor\cdot\rfloor$ is the floor function.
Taking the summation over $k$ as in Eq.~(\ref{general}), we can derive the analytical expression of the fidelity in Eq.~(\ref{fulgraph_fidexact}) from Eq.~(\ref{fulgraph_fidelity2}).

\subsection{Two-dimensional cluster states}

We discuss the estimation of the fidelity of 2D cluster states, which are universal resource states for MBQC.
Here, we focus on rectangular cluster states with $n=q\times r$ ($q,r\in \mathbb N$, $q\neq r$) qubits, because they suffice for universal quantum computation in MBQC.

In contrast to fully-connected graph states, it should be difficult to derive a general expression for the fidelity of 2D cluster states. Meanwhile, the fidelity up to the third order error term can be easily obtained. 
In the case of $q,r>2$, cluster states have no second-order error terms, because they have no stabilizers that consist of two Pauli operators.
As for the third-order error terms, the four generators on the corners of the corresponding rectangular are the only stabilizers that consist of three Pauli operators. 
Thus, the fidelity up to the third-order error can be written as
\begin{align}
F'=(1-p)^n+4(1-p)^{n-3}\left(\frac{p}{3}\right)^3.
\end{align}
The fidelity of the cluster state with $2\times 4$ qubits up to the third-order error is given as
\begin{align}
F'=(1-p)^8+8(1-p)^{5}\left(\frac{p}{3}\right)^3.
\end{align}

\begin{figure}[t]
\includegraphics[width=8cm,clip]{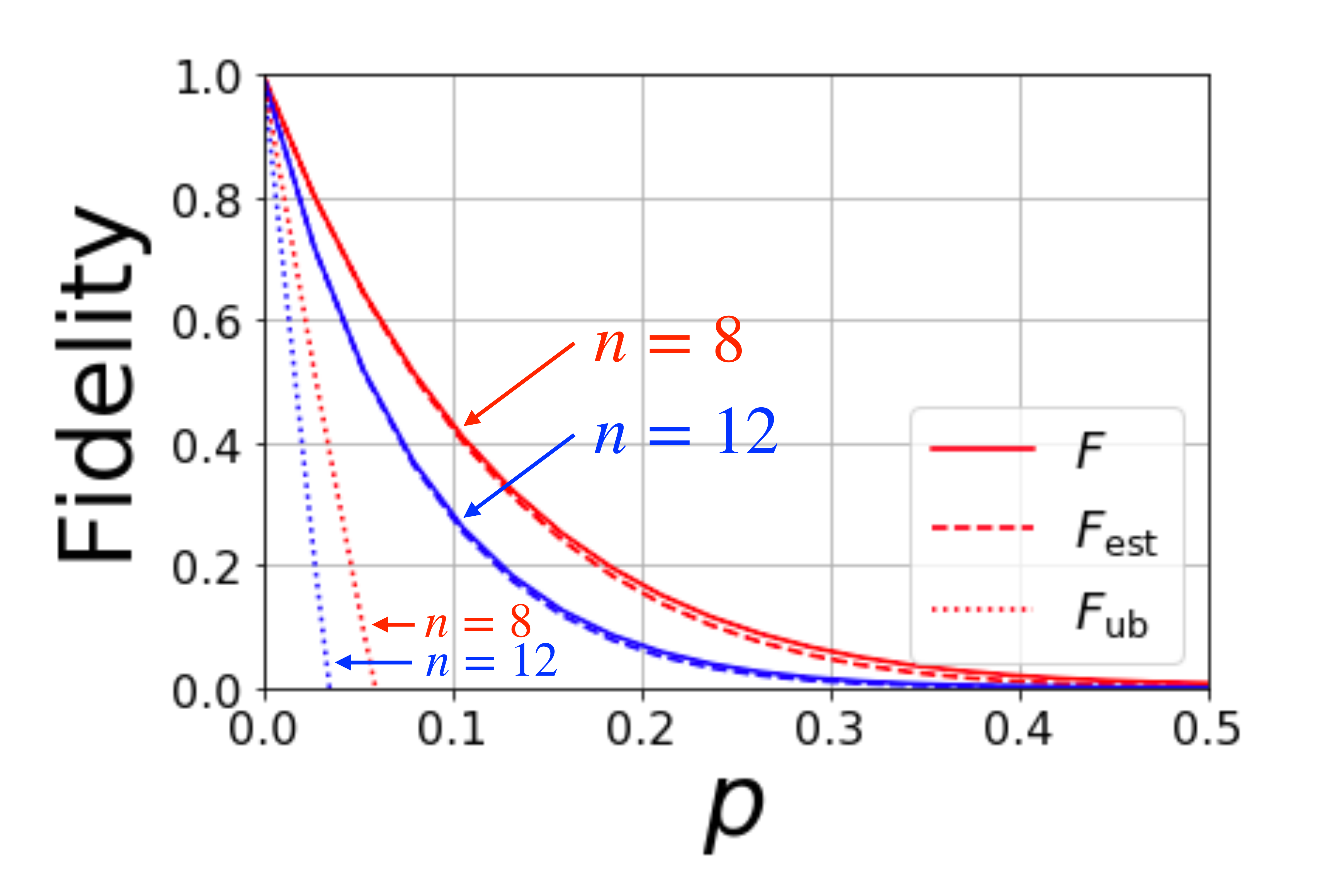}
\caption{Comparison of $F$ (solid lines) and $F_{\rm est}$ (dashed lines) as functions of the error probability $p$ for rectangular cluster states of $2\times 4$ and $3\times 4$ qubits. The dotted lines represent $F_{\rm ub}$ in Eq.~(\ref{Fubcluster}).}
\label{cluster_fidelity}
\end{figure}

Since $F$ has no second-order error terms in both cases, it is expected that $F$ can be well estimated by $F_{\rm est}$. 
Figure~\ref{cluster_fidelity} shows the comparison of the fidelity $F$ and $F_{\rm est}$ for the cluster states with $2\times 4$ and $3\times 4$ qubits. The stabilizers that satisfy the condition $n_I=n/4$ are shown in Figs.~\ref{dverification} (a) and (b). Here, $F$ is evaluated numerically by taking the average of all the stabilizers.
It also demonstrates that the estimation of $F$ by $F_{\rm est}$ improves as $n$ increases.
Compared with Fig.~\ref{fulgraph_fidelity}, the deviation of $F_{\rm est}$ from $F$ is smaller than that for fully-connected graph states.

\subsection{Fidelity estimation for cluster states in the presence of either phase-flip or depolarizing noise}

So far, we have restricted the noise model and fixed $\mathcal{E}$ to the depolarizing channel. 
This restriction can be justified in the case where we can specify the noise model based on knowledge of how the cluster state provided to a verifier is generated in an experiment.
In this subsection, we relax this restriction and consider the possibility of estimating the fidelity of a 2D cluster state by measuring a single stabilizer in the presence of either phase-flip or depolarizing noise. 
It may be useful in the case where the phase-flip or depolarizing noise is randomly applied. It may be also useful in the case where the noise model cannot be decided between the phase-flip and depolarizing noises due to the lack of the knowledge of experimental setups.

We first consider the cluster state of $2\times 4$ qubits. In the presence of the phase-flip error, the fidelity of an $n$-qubit graph state can be estimated by measuring a single stabilizer $S_{\ell}$ that satisfies ${\rm wt}(\ell)=n/2$ \cite{ATYT22}.
Then, the stabilizer 
\begin{align}
S_{\ell}=g_1g_3g_5g_7=X_1Z_2X_3I_4X_5Z_6X_7I_8
\label{stabilizer8}
\end{align}
satisfies both the conditions ${\rm wt}(\ell)=n/2$ and $n_I=n/4$, where the indices of qubits in Eq.~(\ref{stabilizer8}) correspond to those in Fig.~\ref{dverification} (a).
Thus, the fidelity can be estimated by measuring it in the presence of either phase-flip or depolarizing noise.

\begin{figure}[t]
\includegraphics[width=8cm,clip]{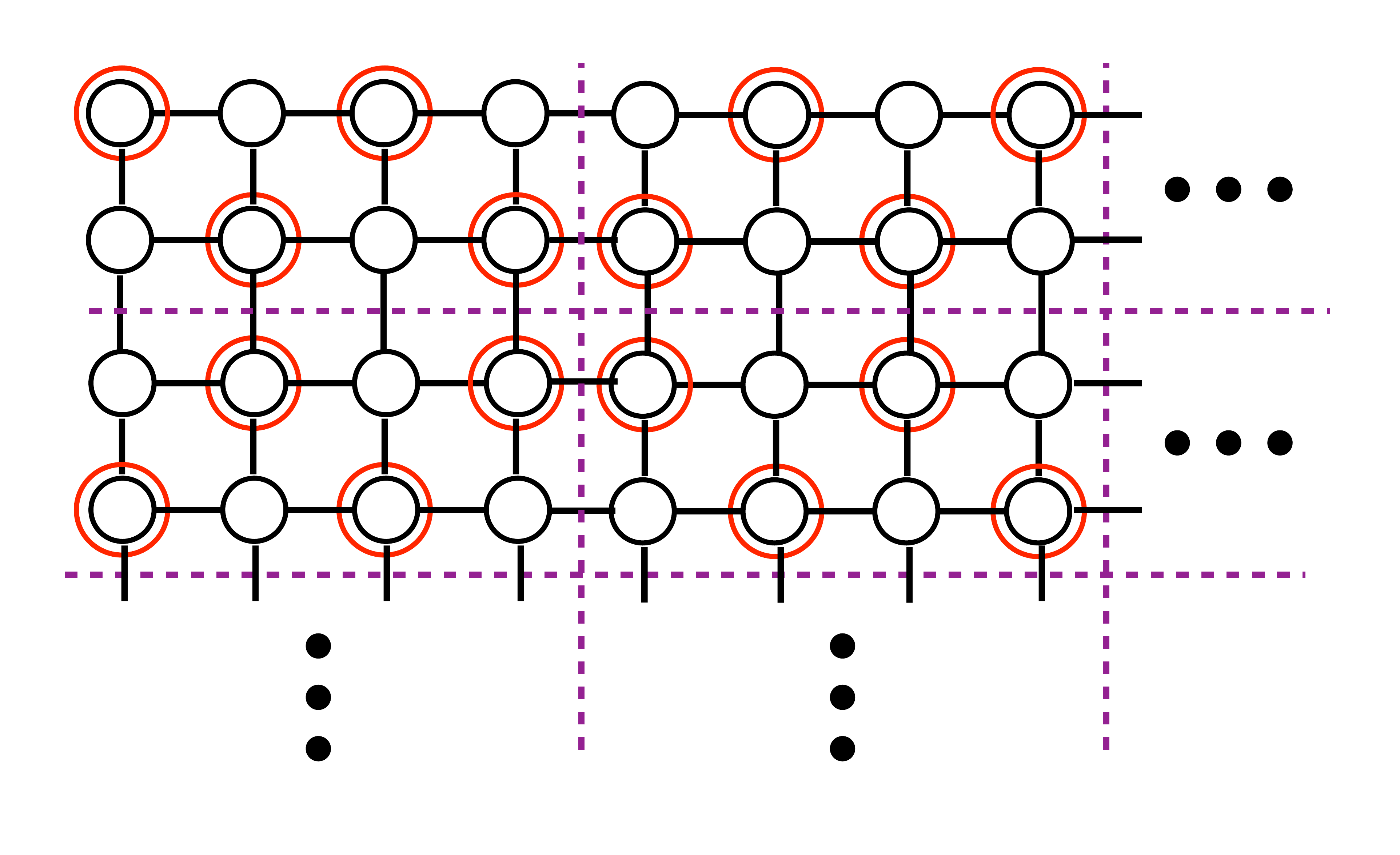}
\caption{Cluster state of $n=4q\times 2r$ qubits. The fidelity of it in the presence of either phase-flip or depolarizing noise can be estimated by a stabilizer that is a product of the generators indicated by the red circles.}
\label{cluster_large}
\end{figure}

Extending the above argument, the fidelity of a large cluster state with $n=4q\times 2r$ qubits can be estimated by measuring a single stabilizer.
For example, choosing generators every $4\times 2$ qubits analogously to Eq.~(\ref{stabilizer8}) as shown in Fig.~\ref{cluster_large}, the stabilizer obtained as their product satisfies both the conditions ${\rm wt}(\ell)=n/2$ and $n_I=n/4$. 

Furthermore, the fidelity of the same cluster state can be estimated by measuring the same stabilizer even in the presence of a more general noise model Eq.~(\ref{generalnoise}) with $p_x=p_y=p/3-\delta$ and $p_z=p/3+2\delta$.
This noise model interpolates the phase-flip and depolarizing noises; it reduces to the depolarizing (phase-flip) noise when $\delta=0$ ($\delta=p/3$). 

From Eq.~(\ref{stblzr_general}), the average of a stabilizer under this noise model can be obtained as
\begin{align}
{\rm Tr}(\rho S_{\ell})
=&
\left(1-\frac{4}{3}p-2\delta\right)^{{\rm wt}(\ell)}\notag\\
&\cdot\left(1-\frac{4}{3}p+4\delta\right)^{n-n_I-{\rm wt}(\ell)},
\label{stblzr_general2}
\end{align}
where we use $n_X+n_Y={\rm wt}(\ell)$ and $n_Z=n-n_I-{\rm wt}(\ell)$.
Thus, setting ${\rm wt}({\ell})=n/2$ and $n_I=n/4$ in Eq.~(\ref{stblzr_general2}), $F_{\rm est}$ for the stabilizer Eq.~(\ref{stabilizer8}) is
\begin{align}
F_{\rm est}(\delta)=
\left(1-\frac{4}{3}p-2\delta\right)^{n/2}
\left(1-\frac{4}{3}p+4\delta\right)^{n/4}.
\label{}
\end{align}
The above expression interpolates $F_{\rm est}=(1-4p/3)^{3n/4}$ for the depolarizing noise ($\delta=0$) and $F_{\rm est}=(1-2p)^{n/2}$ for the phase-flip noise ($\delta=p/3$) \cite{ATYT22}. 

\begin{figure}[t]
\includegraphics[width=8cm,clip]{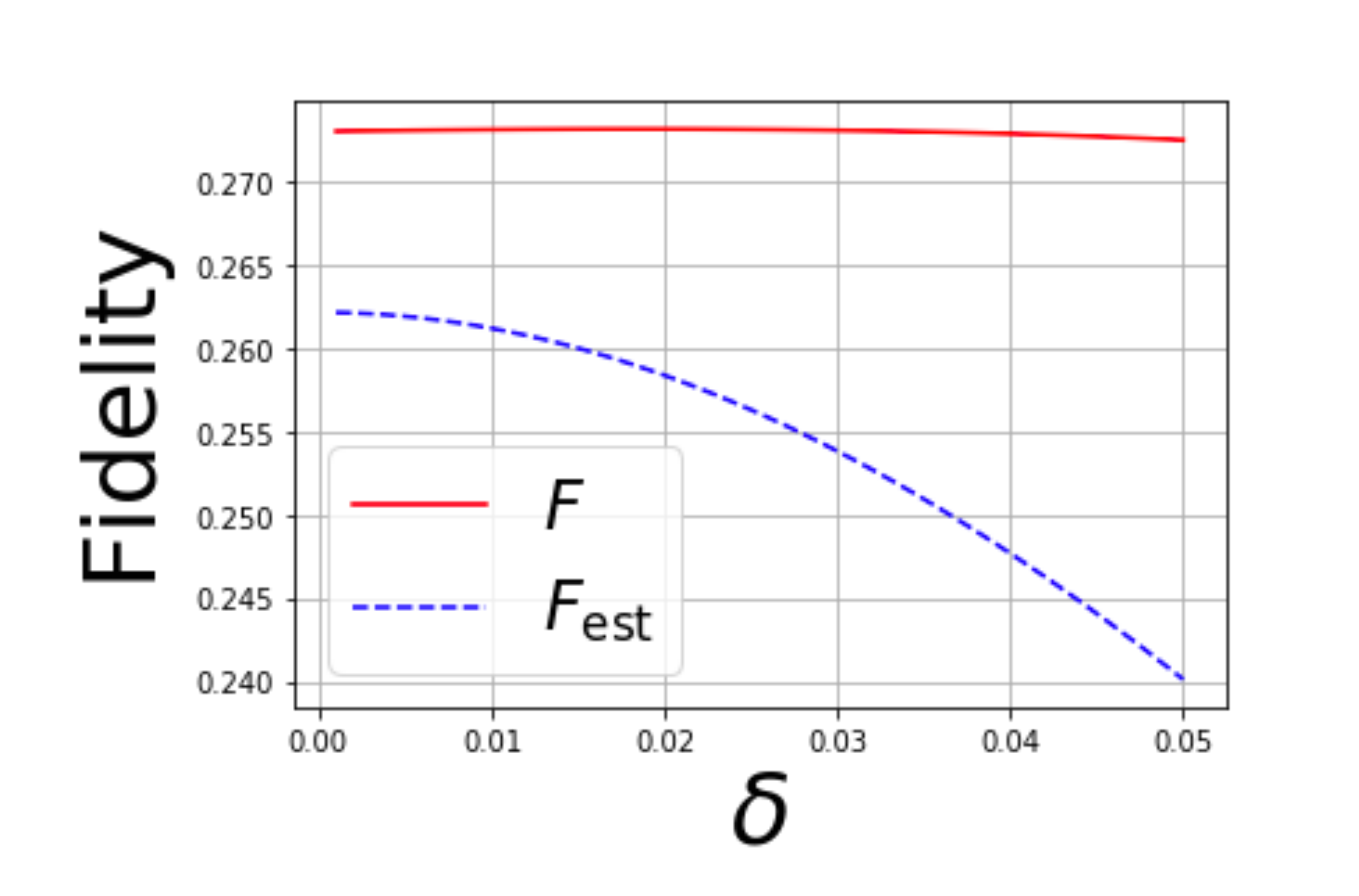}
\caption{Comparison of $F$ and $F_{\rm est}$ for the $(2\times 4)$-qubit cluster state in the presence of the noise model in Eq.~(\ref{generalnoise}) as functions of the parameter $\delta$. We set $p=0.15$.}
\label{fidelity_both}
\end{figure}

Figure~\ref{fidelity_both} shows a comparison of the fidelity $F(\delta)$ and $F_{\rm est}(\delta)$ for the $(2\times 4)$-qubit cluster state in the presence of the noise Eq.~(\ref{generalnoise}) with $p_x=p_y=p/3-\delta$ and $p_z=p/3+2\delta$. From $\partial F_{\rm est}/\partial\delta\leq 0$ in the case of $0\le 2\delta \le 1-4p/3$, $F_{\rm est}(\delta)$ monotonically decreases, as shown in Fig.~\ref{fidelity_both}. When $0\le p\le 1/2$, therefore, $F_{\rm est}(\delta)$ is a lower bound on $F(\delta)$ for any $\delta$ as
\begin{align}
F_{\rm est}(0\leq \delta \leq p/3)\leq F_{\rm est}(\delta=0)=\left(1-\frac{4}{3}p\right)^{3n/4}\notag\\
\leq \tilde F=(1-p)^n<F(\delta),
\end{align}
where we use Theorem \ref{theorem1} in the second inequality. Consequently, a lower bound on the fidelity of a large cluster state with $n=4q\times 2r$ qubits can also be estimated by measuring the stabilizer specified in Fig.~\ref{cluster_large} when $0\le p\le 1/2$.

\medskip
\section{Comparison with previous protocols}
\label{V}

Several verification protocols exist for graph states that work for any type of error~\cite{HM15,MNS16,FH17,TM18,TMMMF19,ZH19,ZH19A,MK20}. The lower bound of the fidelity obtained in some of them becomes loose in general.
It has been shown that the necessary number of measurement settings can be improved to $n$ from $2^n$ by using the union bound \cite{TM18}, where the obtained lower bound of the fidelity is given as
\begin{eqnarray}
F_{\rm ub}=1-\sum_{i=1}^n\left\{1-{\rm Tr}\left[\rho\left(\cfrac{I^{\otimes n}+g_i}{2}\right)\right]\right\}
\end{eqnarray}
in the limit of large $N$.

For any $n$-qubit fully-connected graph state, using ${\rm Tr}(\rho g_i)=(1-4p/3)^n$, the lower bound $F_{\rm ub}$ is calculated as
\begin{align}
F_{\rm ub}&=1-\frac{n}{2}+\frac{n}{2}\left(1-\frac{4}{3}p\right)^n
\label{Fubfully}\\
&\simeq 1-\frac{2}{3}n^2 p\quad(p\ll 1).
\end{align}
For any $n=q\times r$ ($q,r\geq 2$)-qubit cluster state, the lower bound $F_{\rm ub}$ is calculated as
\begin{align}
&F_{\rm ub}=1-\frac{n}{2}+\frac{1}{2}\left[2(q+r-4)\left(1-\frac{4}{3}p\right)^4\right.\notag\\
&\left.+4\left(1-\frac{4}{3}p\right)^3+(q-2)(r-2)\left(1-\frac{4}{3}p\right)^5\right]
\label{Fubcluster}\\
&\simeq 1-\frac{2}{3}\left[5n-2(q+r)\right]p\quad(p\ll 1).
\end{align}
$F_{\rm ub}$ becomes loose as $n$ increases in both types of graph states, which is in sharp contrast with the fact that $F_{\rm est}$ becomes tight as $n$ increasing.
Figures~\ref{fulgraph_fidelity} and \ref{cluster_fidelity} show that our estimated value $F_{\rm est}$ is close to the true value $F=\langle G|\rho|G\rangle$ even when the number $n$ of qubits and $p$ are large. Meanwhile, the deviation of $F_{\rm ub}$ from $F$ increases as $p$ increasing as shown in Figs.~\ref{fulgraph_fidelity} and \ref{cluster_fidelity}.

\medskip
\section{Conclusion and discussion
\label{VI}}

We have proposed a verification protocol for graph states assuming the depolarizing channel.
A remarkable feature of our verification protocol is that the fidelity of an $n$-qubit graph state can be estimated by just measuring a single stabilizer $S_{\ell}$ that satisfies the condition $n_I=n/4$, where $n_I$ denotes the number of identity operators in $S_{\ell}$, and so it requires only one measurement setting.
Furthermore, we have shown that the estimation improves as the number $n$ of qubits increases. 
We have also derived a simple analytic expression for the average of a stabilizer in the depolarizing channel.
We have applied our protocol to fully-connected graph states as well as cluster states and have demonstrated its usefulness.
Furthermore, we have evaluated our protocol for other noise models other than the depolarizing noise and have compared it with previous protocols.
Our protocol should be useful even when it is unknown whether the actual physical noise model is the phase-flip or depolarizing noise.

We note the scope of applications of our protocol in the presence of the depolarizing noise.
Our protocol, which estimates the fidelity of the graph state under the depolarizing noise, is specifically useful in the situation where we have almost no information about the actual physical noise channel for the following reason.
For a single-qubit state, the depolarizing noise introduces a probabilistic mixture of the ideal state and the maximally mixed state with probability $4p/3$ (see Eq.~(\ref{dplns})).
From Ref.~\cite{W11}, this noise model is valuable when knowledge about the underlying noise channel is quite limited.
In essence, the depolarizing noise is used as a representative model due to the lack of detailed information about the specific noise model.

Introducing the step of estimating the parameter $p$ could indeed be an approach.
In the approach, to obtain an estimated value of the fidelity, Eq.~(\ref{fidelity_graph}) has to be calculated with the estimated value of $p$, which increases the required time for the estimation.
On the other hand, our protocol can directly estimate the fidelity without calculating Eq.~(\ref{fidelity_graph}).

As a future work, it would be interesting to extend our results to correlated noises such as the global depolarizing noise
\begin{eqnarray}
\mathcal{E}(\rho)=(1-p)\rho+p\cfrac{I^{\otimes n}}{2^n},
\end{eqnarray}
where $\rho$ is any $n$-qubit state, and $p$ is the error probability.
Since this error model is observed in actual experiments such as the quantum supremacy experiment in Ref.~\cite{google}, by doing so, we can make our results more practical.

\medskip
\section*{ACKNOWLEDGMENTS}
We thank anonymous reviewers for insightful comments. ST is supported by the Japan Society for the Promotion of Science Grant-in-Aid for Scientific Research (KAKENHI Grant No.~19K03691).
RY is supported by JSPS Grant-in-Aid for Scientific Research (KAKENHI Grant No.~19K14616 and 20H01838).
YT is supported by the MEXT Quantum Leap Flagship Program (MEXT Q-LEAP) Grant Number JPMXS0118067394 and JPMXS0120319794, JST [Moonshot R\&D -- MILLENNIA Program] Grant Number JPMJMS2061, and the Grant-in-Aid for Scientific Research (A) No.JP22H00522 of JSPS. SY is supported by Grant-in-Aid for JSPS Fellows (Grant No.\,JP22J22306).

\medskip

\section*{APPENDIX A: Proof of Lemma \ref{lemma1}}
\label{A}

\begin{proof}
Using Eq.~(\ref{graph}), the expectation value $\langle G|(\prod_{k=1}^m\sigma_{\mu_ki_k})|G\rangle$
can be written as
\begin{align}
&\langle G|\left(\prod_{k=1}^m\sigma_{\mu_ki_k}\right)|G\rangle\notag\\
&=\langle+|^{\otimes n}U_{CZ}^\dagger\left(\prod_{k=1}^m\sigma_{\mu_ki_k}\right)U_{CZ}|+\rangle^{\otimes n},
\label{PauliG}
\end{align}
where $U_{CZ}\equiv\prod_{e\in E}CZ_e$.
Given that $CZ$ is a Clifford operator, $U_{CZ}^\dagger(\prod_{k=1}^m\sigma_{\mu_ki_k})U_{CZ}$
is also a tensor product of Pauli operators with a sign $+$ or $-$. 
A tensor product of Pauli operators averaged by $|+\rangle^{\otimes n}$ yields $\pm 1$ if it is a tensor product of $X$ and/or $I$ and zero otherwise.
Thus we obtain $\langle G|(\prod_{k=1}^m\sigma_{\mu_ki_k})|G\rangle^2=$ $0$ or $1$.
For $\langle G|(\prod_{k=1}^m\sigma_{\mu_ki_k})|G\rangle$
being nonzero, 
there must exist $s\in\{+1,-1\}$ and the set $A\subseteq\{1,2,\ldots,n\}$ such that
\begin{align}
&U_{CZ}^\dagger\left(\prod_{k=1}^m\sigma_{\mu_ki_k}\right)U_{CZ}=s \left(\prod_{i\in A} X_i\prod_{j\in\bar{A}} I_j\right),\\
&\therefore s\left(\prod_{k=1}^m\sigma_{\mu_ki_k}\right)=U_{CZ}\left(\prod_{i\in A} X_i\prod_{j\in\bar{A}} I_j\right)U_{CZ}^\dagger,
\end{align}
where $\bar{A}$ is the complement of $A$.
When $\langle G|(\prod_{k=1}^m\sigma_{\mu_ki_k})|G\rangle$
is nonzero, thus,
$\prod_{k=1}^m\sigma_{\mu_ki_k}$ or
$-\prod_{k=1}^m\sigma_{\mu_ki_k}$ coincides with one of the stabilizers of $|G\rangle$. 
\end{proof}

\section*{APPENDIX B: Derivation of Eq.~\eqref{stblzr_general} }
\label{B}

In this Appendix, we derive Eq.~(\ref{stblzr_general}) and provide with an alternative derivation of Eq.~\eqref{generalshort}. 
For convenience, we express the general noise model Eq.~(\ref{generalnoise}) as 
\begin{equation}
\mathcal{E}(\cdot )=\sum_{\mu=0,1,2,3} p_\mu \sigma_\mu (\cdot) \sigma_\mu, 
\label{generalnoise2}
\end{equation}
where $p_1=p_x$, $p_2=p_y$, $p_3=p_z$, and $p_0=1-p_x-p_y-p_z$, $\sigma_0=I$, $\sigma_1=X$, $\sigma_2=Y$, and $\sigma_3=Z$.
\par
Using the cyclic property of trace and the anti-commuting property of the Pauli matrices, we can derive the following relation:
\begin{align}
&\mathrm{Tr}\left[\sigma_\nu \mathcal{E}(\cdot )\right]
=\sum_{\mu=0,1,2,3}p_\mu\mathrm{Tr} \left[ \sigma_\mu \sigma_\nu \sigma_\mu (\cdot) \right]\notag\\
&=f_\nu \mathrm{Tr} \left[\sigma_\nu (\cdot)\right],\quad (\nu=0,1,2,3).
\end{align}
Here, $f_\nu$ is given as
\begin{align}
f_\nu=p_0+\sum_{\mu=1,2,3}p_\mu (-1)^{1+\delta_{\mu,\nu}}
\end{align}
in the case of $\nu=1,2,3$, and $f_0=1$.
Using the above relation, we obtain 
\begin{align}
&\mathrm{Tr}\left[\left(\bigotimes_{i=1}^n \tau_i \right)\mathcal{E}^{\otimes n}(\cdot )\right]\notag\\
&=\mathrm{Tr}_1\cdots \mathrm{Tr}_n\left[\left(\bigotimes_{i=1}^n \tau_i \right) \mathcal{E}^{\otimes n}(\cdot )\right]\notag\\
&=f_1^{n_X}f_2^{n_Y}f_3^{n_Z}\mathrm{Tr}\left[\left(\bigotimes_{i=1}^n \tau_i \right) (\cdot )\right].
\end{align} 
Thus, Eq.~(\ref{stblzr_general}) can be obtained as
\begin{align}
\mathrm{Tr}\left(S_{\ell} \rho\right) =& \mathrm{Tr}\left[\left((-1)^s\bigotimes_{i=1}^n \tau_i \right)\mathcal{E}^{\otimes n}(|G\rangle\langle G| )\right]\notag\\
=&f_1^{n_X}f_2^{n_Y}f_3^{n_Z}\nonumber\\
&\times\mathrm{Tr}\left[\left((-1)^s \bigotimes_{i=1}^n  \tau_i \right) (|G\rangle\langle G| )\right]\notag\\
=&f_1^{n_X}f_2^{n_Y}f_3^{n_Z}\langle G|S_{\ell} |G\rangle \nonumber\\
=&\left(1-2p_y-2p_z\right)^{n_X}\left(1-2p_z-2p_x\right)^{n_Y}\notag\\
&\times\left(1-2p_x-2p_y\right)^{n_Z}.
\end{align}
For the depolarizing noise, setting $p_x=p_y=p_z=p/3$, we obtain 
\begin{align}
\mathrm{Tr}\left(S_{\ell} \rho\right) =\left(1-\frac{4}{3}p\right)^{n-n_I},   
\end{align} 
where we use $n_X+n_Y+n_Z=n-n_I$.

\section*{APPENDIX C: Proof of Theorem \ref{theorem1}}
\label{C}

\begin{proof}
Let $X\equiv (1-p)^4$ and $Y\equiv (1-4p/3)^3$, then $\tilde F=X^k$ and $F_{\rm est}=Y^k$. 
Using $X\ge Y$ for $p\in[0,1]$, and $X-Y=[(p-22/27)^2+2/729]p^2$,
\begin{align}
0\leq &\tilde F-F_{\rm est}=X^k-Y^k\notag\\
=&(X-Y)(X^{k-1}+X^{k-2}Y+\dots +XY^{k-2}+Y^{k-1})\notag\\
\le& k(X-Y)X^{k-1}\notag\\
\le& kp^2\left(-\frac{1}{2}p+\frac{2}{3}\right)(1-p)^{4(k-1)}\label{polynomialp}\\
\le& kp_0^2\left(-\frac{1}{2}p_0+\frac{2}{3}\right)(1-p_0)^{4(k-1)}\le \frac{2}{3}kp_0^2<\frac{2}{3k}.\label{inequalityk}
\end{align}
Here, the right-hand-side of Eq.~(\ref{polynomialp}) takes its maximum value at $p_0$, which satisfies 
\begin{equation}
    p_0=\frac{16k+1-\sqrt{256k^2-352k+97}}{6(4k-1)}.
\end{equation}
The last inequality in Eq.~(\ref{inequalityk}) can be obtained from the inequality $p_0<1/k$.
\end{proof}

\section*{APPENDIX D: Derivation of Eq.~\eqref{uprbnd_GHZ} }
\label{D}

$F-F_{\rm est}$ for the fully-connected graph states with $n=8k$ qubits can be written as
\begin{align}
&F-F_{\rm est}=\frac{1}{2}\left\{\left[\left(1-\frac{2}{3}p\right)^n-(1-p)^n\right]\right.\notag\\
&\left.-\left[(1-p)^n-\left(1-\frac{4}{3}p\right)^n\right]\right\}\notag\\
&+\frac{1}{2}\left(\frac{2}{3}p\right)^n+\left[(1-p)^n-\left(1-\frac{4}{3}p\right)^{6k}\right].
\label{deltaF}
\end{align}
Note that $F-F_{\rm est}\geq 0$ holds because $F\geq \tilde F$ from Eq.~(\ref{fidelity_graph}), and $\tilde F\geq F_{\rm est}$ from Eq.~(\ref{lwrbnd}). From the inequalities 
\begin{align}
&\left(\frac{2}{3}p\right)^n\leq \left(\frac{2}{3}\right)^n,
\label{inequality_GHZ1}\\
&(1-p)^n-\left(1-\frac{4}{3}p\right)^{6k}< \frac{1}{3k},\label{inequality_GHZ2}
\end{align}
it is necessary to evaluate the upper bound of the terms inside the bracket $\{\cdot\}$ on the right hand side of Eq.~(\ref{deltaF}). Note that Eq.~(\ref{inequality_GHZ2}) holds due to Theorem~\ref{theorem1}. Let $s\equiv 1-p$, $t\equiv 1-(2/3)p$, and $u\equiv 1-(4/3)p$. For $0\leq p\leq 3/4$, it can be evaluated as
\begin{widetext}
\begin{align}
(t^n-s^n)-(s^n-u^n)&=(t-s)(t^{n-1}+t^{n-2}s+\dots+s^{n-1})-(s-u)(s^{n-1}+s^{n-2}u+\dots+u^{n-1})\notag\\
&\le (t-s) nt^{n-1}-(s-u)ns^{n-1}= \frac{p}{3}n(t^{n-1}-s^{n-1})\notag\\
&=\frac{p}{3}n(t-s)(t^{n-2}+t^{n-3}s+\dots+ts^{n-3}+s^{n-2})\notag\\
&\le \frac{n(n-1)}{9}p^2t^{n-2}=\frac{n(n-1)}{9}p^2\left(1-\frac{2}{3}p\right)^{n-2}\leq \left(1-\frac{1}{n}\right)\left(1-\frac{2}{n}\right)^{n-2}.
\label{inequality_GHZ3}
\end{align}
\end{widetext}
We use the relation $(t^{n-1}-t^{n-i}s^{i-1})-(s^{n-1}-s^{n-i}u^{i-1})\geq 0$ for integer $i\in[1,n]$, which can be easily proved by induction, for the first inequality.
We also use the fact that $p^2(1-(2/3)p)^{n-2}$ takes the maximum value at $p=3/n$ for $0\leq p\leq 3/4$ for the final inequality. Combining Eqs.~(\ref{inequality_GHZ1}), (\ref{inequality_GHZ2}), and (\ref{inequality_GHZ3}), we obtain Eq.~(\ref{uprbnd_GHZ}). 
The right hand side of Eq~(\ref{inequality_GHZ3}) converges to $1/e^2$ due to the relation
\begin{equation}
\lim_{n\to\infty}\left(1-\frac{2}{n}\right)^n=\frac{1}{e^2}.
\end{equation}
Therefore, the upper bound in Eq.~(\ref{uprbnd_GHZ}) converges to $1/(2e^2)$.

\end{document}